\documentclass{article}
\usepackage{amsmath, amssymb, graphics, setspace}

\newcommand{\mathsym}[1]{{}}
\newcommand{\unicode}[1]{{}}

\begin{document}

\title{First-principles lattice-gas Hamiltonian revisited: O-Pd(100) { }}
\author{Wolfgang Kappus\\
\\
wolfgang.kappus@t-online.de}
\date{v3: 2017-11-30}
\maketitle

\section*{Abstract}

The methodology of deriving an adatom lattice-gas Hamiltonian (LGH) from first principles (FP) calculations is revisited. Such LGH cluster expansions
compute a large set of lateral pair-, trio-, quarto interactions by solving a set of linear equations modelling regular adatom configurations and
their FP energies. The basic assumption of truncating interaction terms beyond fifth nearest neighbors does not hold when adatoms show longer range
interactions, e.g. substrate mediated elastic interactions. O-Pd(100) as a popular reference is used to propose a long range elastic interaction
alternative including many-body trio- and quarto terms with just 3 parameters fitted to FP calculations. A key feature of the interaction alternative
is its analytic nature, allowing statistical methods different from Monte Carlo simulations to derive surface order. The assumptions made are discussed
and ways to further verify and apply the model are outlined. 

\section*{1. Introduction}

Interactions of adatoms are a subject of continuous interest, various different interaction mechanisms have been described in detail [1]. Adatom
structures like superlattices, nanodot arrays, nanostripes, strain relief patterns are interesting for various general and technological reasons;
reviews were given in [2,3,4,5]. Lateral interactions also govern the ordering behavior of adatoms and thus the catalytic activities of surfaces
[6]. Lattice models are the discrete representations of surface properties and usable for predicting structure and phase diagrams of adatom configurations
[1].

Thanks to computational power and advanced algorithms the lateral interactions of adatoms have been determined from first principles using density-functional
theory (DFT) and used as inputs to derive lattice-gas Hamiltonians (LGHs) with cluster expansion (CE) techniques [7]. The accuracy of this method
was analyzed for the O-Pd(100) system and its results were applied to model surface ordering of atomic adsorbates at that system [8]. These first
principle calculations determined the location of O adatoms preferably on fourfold coordinated surface sites. The subsequent lattice gas expansion
postulated many trio- and quarto interaction parameters to fit the DFT results. Those calculations focused on the range up to about half coverage
since experiments showed the onset of surface reconstruction when the coverage is increased beyond half coverage [9]. { } { } 

Key assumption in those calculations were vanishing interactions from 6th next neighbors, allowing to derive short range lateral interaction energies
from ordered adatom configurations with systems of linear equations.

Since similar calculations of the O-Ag(111) system showed significant lateral displacements [10] the question is raised in the following, what impact
longer range interactions may have also in the O-Pd(100) system. An \(s^{-3}\) type interaction would lead to an \(s^{-1}\) dependency in lattice
sums and show results quite different from short range interactions.

In the following sections the DFT results of [8] for the O-Pd(100) system will be interpreted differently in terms of elastic adatom interactions
and an appropriate lattice gas Hamiltonian will be derived. To facilitate the comparison of the two approaches the terminology of [8] will be used
in parts.

Strain mediated adatom interactions with the basic \textup{ \(s^{-3}\)} distance proportionality and the strong influence of the substrates elastic
anisotropy were discussed already in the 1970 decade based on elastic continuum theory [11,12,13] . Usage of an elastic mode theory [14] allowed
to handle the \textup{ \(s^{-3}\)} singularity at the origin [12]. A review of elastic effects on surface physics was given in [15]. { } 

The results of [8] for the O-Pd(100) system claim the relevance of trio- and quarto interactions; this caused an extension of the pair interaction
theory of [12]. The presentation of such extension would overload this analysis and therefore is given elsewhere [16]. 

In the pair interaction model of [12] the interaction of adatoms is mediated by their strain fields generated by single adatoms exerting isotropic
stress to their vicinity. Isotropic stress is a consequence of adatom locations on sites with 4-fold symmetry. In a lattice description adatoms would
exert forces to their immediate substrate neighbors creating a displacement field equivalent to a strain field.

In other cases where adatoms interact directly e.g. via their dipole moment or via electronic overlap, pairs of adatoms may create substrate strain
{ }by stretching or compressing the substrate to balance the forces. The strain field of such pairs will mediate interactions between pairs and also
between pairs and monomers, an interpretation of trio and quarto interactions. If forces between dimer constituents are central, the model can be
kept simple and the resulting stress field can (with symmetry restrictions) be described with one more free parameter [16] than the previous pair
interaction model [12]. { }

Since within [8] DFT-calculated energies for only a few ordered adatom configurations have been published, the comparison between the elastic model
and the DFT results must stay incomplete for the time being.

The elastic model is analytic and allows differentiation. This allows in a forthcoming analysis to calculate adatom pair distributions to evaluate
the near order of adsorbates.

The paper is organized as follows:\\
After outlining the general motivation in section 1, basic interaction model details are recalled in section 2. In section 3 model parameters are
fitted to the first principles calculations of [8]. In section 4 applicability and limitations of the model are discussed and open questions are
addressed. Section 5 closes with a summary of the results.

\section*{2. Adatom Interaction Model}

In this section the lattice-gas Hamiltonian is employed in which the total free binding energy of one component configurations are expanded in sums
of discrete interactions between lattice sites. These discrete interactions are related to the energy of configurations calculated by first principles
methods as done for O-Pd(100) in [8].

\subsection*{2.1. Lattice-gas Hamiltonian}

The total free binding energy H is assumed to be a sum of site energies \(F_b^{\text{on}-\text{site}}\), pair interactions \(V_{\text{\textit{ij}}}\),
trio interactions \(V_{\text{\textit{ijk}}}\) and higher terms { } { } { } 

\[H=F_b^{\text{on}-\text{site}}\sum _i n_i + \sum _i \sum _{j<i} V_{\text{ij}}n_i n_j +\sum _i \sum _{j<i} \sum _{k<j<i} V_{\text{ijk}}n_i n_jn_k
+\text{  }\text{...}\text{                                       }(2.1)\]

with occupation numbers \(n_i\). The discrete interactions will not individually be derived from DFT calculations as in [8] but taken from a substrate
strain mediated interaction model outlined in the following section.  { } 

\subsection*{2.2. Elastic interaction model }

The occurrence of trio- and quarto interactions in [8] causes an extension of current elastic pair interaction models. While elastic pair interactions
are related to the stress fields of single adatoms, adatom dimers will generate elastic trio- and quarto interactions in the following model, a condensed
version of [16]. { } 

Adatoms or dimers exert forces on their substrate neighbors leading to a displacement of substrate atoms to balance those forces. Such displacements
will increase or decrease the energy of neighboring adatoms or dimers. In a continuum description adatoms and dimers exert stress parallel to the
surface leading to substrate strain which in turn can lead to an attraction or a repulsion of neighboring adatoms or dimers. It will be shown that
the strength of such strain mediated interaction will depend on the stress adatoms or dimers exert on the substrate and on the stiffness of the substrate.

The elastic energy of a substrate with adatoms in a continuous description is given by the sum of two parts, the energy of the distorted substrate
and the energy of adatoms exerting tangential forces on the substrate { }

\[H_{\text{el}} = \frac{1}{2} \int _V\pmb{\text{\textit{$\epsilon (r)$}}\text{\textit{$ $}}c\text{\textit{$ $}}\text{\textit{$\epsilon (r)$}}\text{\textit{$dr$}}}+\text{
  }\int _S\pmb{\text{\textit{$\epsilon (s)$}}}\pmb{\text{\textit{$\pi (s)$}}}\pmb{\text{\textit{$ds$}}} .\text{                                 
    }(2.2)\]

Here \pmb{ \textit{ $\epsilon $ }}= [\(\epsilon _{\alpha \beta }\)] denotes the strain tensor field, \pmb{ \textit{ c }}= [\(c_{\alpha \beta \mu
\nu }\)] denotes the elastic constants tensor, and \pmb{ \textit{ $\pi $ }}= [\(\pi _{\mu \nu }\)] denotes the force dipole- or stress tensor field.
The integrals comprise the bulk V or the surface S. The strain field \pmb{ \textit{ $\epsilon $(r)}} is related to the displacement field \pmb{ \textit{
u(r)}} by

\[\epsilon _{\alpha \beta }(\pmb{r})=\frac{1}{2} \left(\nabla _{\alpha }u_{\beta }(\pmb{r})+\nabla _{\beta } u_{\alpha }(\pmb{r})\right) .\text{
                                        }(2.3)\]

Following [16] the stress field \pmb{ \textit{ $\pi $(s)}} is superimposed of \textit{ np} different types \pmb{ \textit{ \(\pi _k\)}}\pmb{ \textit{
(s)}}

\[\pmb{\pi \text{\textit{$($}}s\text{\textit{$)$}}}= \sum _{k=1}^{\text{\textit{np}}}  \pmb{\text{\textit{$\pi _k$}}\text{\textit{$($}}s\text{\textit{$)$}}}=\sum
_{k=1}^{\text{\textit{np}}}  \pmb{\text{\textit{$P_k$}}}\rho _k(\pmb{s})\text{  },\text{                                          }(2.4)\]

where we have introduced \textit{ np} different types of { }stress tensors \(\pmb{\text{\textit{$P_k$}}}\) and of adatom monomer or dimer densities
\(\rho _k(\pmb{s})\). \\
On (100) surfaces with adatom positions of fourfold coordinated sites 

\(\pmb{\text{\textit{$P_1$}}}\)=\(P_1\)[\(\delta _{\alpha \beta }\)] stand for an isotropic monomer stress tensor (or force dipole tensor), 

\(\pmb{\text{\textit{$P_2$}}}\)=\(P_2\)[\(\delta _{\text{$\alpha $1}}\)\(\delta _{\text{$\beta $1}}\)] and \(\pmb{\text{\textit{$P_3$}}}\)=\(P_3\)[\(\delta
_{\text{$\alpha $2}}\)\(\delta _{\text{$\beta $2}}\)] stand for anisotropic dimer stress tensors. 

\(\rho _1(\pmb{s})\) then stands for the adatom monomer density distribution, \(\rho _2(\pmb{s})\) for the x-directed adatom dimer density distribution
and \(\rho _3(\pmb{s})\) for the y-directed adatom dimer density distribution. \\
With Eq. (2.4) Eq. (2.2) reads

\[H_{\text{el}} = \frac{1}{2}\text{  }\int _V \pmb{\text{\textit{$\epsilon (r)$}}\text{\textit{$ $}}c\text{\textit{$ $}}\text{\textit{$\epsilon (r)$}}\text{\textit{$dr$}}}\underline{
}+ \sum _{k=1}^{\text{np}} \int _S \pmb{\text{\textit{$\epsilon (s)$}}} \pmb{\text{\textit{$P_k$}}}\rho _k(\pmb{s}) \pmb{\text{\textit{$ds$}}}\text{
 }.\text{                                                }(2.5)\]

The strain field \pmb{ \textit{ $\epsilon $(r)}} is determined for given densities \(\rho _k(\pmb{s})\) by the requirement of mechanical equilibrium

\[\delta  \left.H_{\text{el}}\right/\delta  u_{\alpha }\pmb{\text{\textit{$($}}}\pmb{\text{\textit{$ $}}}\pmb{r})\underline{ } = 0 .\text{      
                                                              }(2.6)\]

After an expansion in plane waves and calculations beyond the scope of this paper we end up at [16] 

\[H_{\text{el}}=\frac{1}{2} \sum _{k=1}^{\text{np}} \sum _{l=1}^{\text{np}}  \int _S \int _S \rho _k(\pmb{s}) V_{\text{\textit{kl}}}\left(\pmb{s},\pmb{\text{\textit{$s'$}}}\right)
\rho _l\left(\pmb{\pmb{s}'}\right) \pmb{\text{\textit{$ds$}}} \pmb{\text{\textit{$ds'$}}}\pmb{\text{\textit{$,$}}}\text{                        
       }(2.7)\text{                         }\]

where the elastic interaction between adatoms (or dimers) of type \textit{ k} at \pmb{ \textit{ s}} and adatoms (or dimers) of type \textit{ l} at
\(\pmb{\text{\textit{$s'$}}}\) \(V_{kl}\) (\pmb{ \textit{ s}},\pmb{ \textit{ s{'}}}) will be given in Eq. (2.14).

\subsection*{2.3. Eigenvalue equation and elastic interaction }

Recalling [16] the elastic interaction \(V_{kl}\) (\pmb{ \textit{ s}},\pmb{ \textit{ s{'}}}) results from expanding the displacement field, given
by freedom of stress in the bulk and freedom of stress normal to the surface, in plane waves. The vector 

\[\left.\pmb{g}_m\pmb{\text{\textit{$($}}}\pmb{\kappa }\right)= i\pmb{\kappa }+\pmb{n}q_m\text{                                                 
                                  }(2.8)\]

denotes the sum of a wave vector \pmb{ \textit{ $\kappa $ }} parallel to the surface and a normal component. The normal vector \pmb{ \textit{ n}}
should not be confused with the occupation numbers used in Eq. (2.1). 

Freedom of stress in the bulk becomes 

\[\left[c_{\alpha \beta \mu \nu }g_{\beta }^{\text{\textit{$m$}}}g_{\mu }^{\text{\textit{$m$}}}\right]a_{\nu }^{\text{\textit{$m$}}}=0,\text{   
         }(2.9)\]

where the \(\pmb{\text{\textit{$a_m$}}}\) are normalized eigenvectors. The label \textit{ m} denotes 3 displacement modes and the \(q_m\) are roots
of the characteristic polynomial in \textit{ q} in Eq. (2.9). In this section \textit{ m} appears as lower index in vectors and as upper index in
vector elements. Freedom of stress normal to the surface reads

\[\left[c_{\alpha \beta \mu \nu }n_{\beta }g_{\mu }^{\text{\textit{$m$}}}a_{\nu }^{\text{\textit{$m$}}}-\omega _{\text{\textit{kl}}}^{-1}(\pmb{\kappa
})P_{\alpha \beta }^{\text{\textit{$k$}}}P_{\mu \nu }^{\text{\textit{$l$}}}g_{\beta }^{\text{\textit{$m$}}}g_{\mu }^{\text{\textit{$m$}}}a_{\nu }^{\text{\textit{$m$}}}\right]A^{\text{\textit{$m,\text{kl}$}}}=0,\text{
            }(2.10)\]

where the \(\omega _{\text{\textit{kl}}}(\pmb{\kappa })\) are eigenvalues of the linear equations for all combinations of stress tensors \(\pmb{\text{\textit{$P_k$}}}\)=\(\left[P_{\alpha
\beta }^k\right.\)]. For further numerical treatment the eigenvalues are expanded in a cosine series

\[\omega _{\text{\textit{kl}}}(\pmb{\kappa })=\left| \pmb{\kappa }\right| \sum _{p=0}^{\text{\textit{pmax}}} \omega _{\text{\textit{kl}},p} \cos
(p \phi ) .\text{                                                                        }(2.11)\]

\(p\) attains the values 0, 4, 8 in the monomer case and 0, 2, 4 in the dimer case. $\phi $ is the angle between the x- and the \pmb{ \textit{ $\kappa
$}} direction. The \(\omega _{\text{\textit{kl}},p}\) are proportional to the product of stress parameters \(P_k\)\(P_l\) and inverse proportional
to the elastic constant \(c_{44}\) defining a dimensionless constant { }\(\hat{\omega }_{\text{\textit{kl}}\text{\textit{$,$}}p}\)

\[\omega _{\text{\textit{kl}},p}\text{  }=\text{  }\hat{\omega }_{\text{\textit{kl}}\text{\textit{$,$}}p}P_kP_l/c_{44}\text{  }.\text{          
                                                                      }(2.12)\text{                    }\]

The elastic interaction \(V_{kl}\) (\pmb{ \textit{ s}},\pmb{ \textit{ s{'}}}) { }in Eq. (2.10) { }is given by [16]

\[V_{\text{\textit{kl}}}\left(\pmb{s},\pmb{\text{\textit{$s'$}}}\right)=(2 \pi )^{-1} \sum _{p=0}^{\text{\textit{pmax}}} \omega _{\text{\textit{kl}},p}
\cos (p \chi ) \cos (p \pi /2)\int _0^{\kappa _B}\kappa ^2J_p(\kappa  s)d\kappa \text{  },\text{                                     }(2.13)\text{
 }\]

where \(J_p\) are Bessel functions of the order \textit{ p, s}=Abs(\pmb{ \textit{ s}}-\(\pmb{\text{\textit{$\pmb{s}'$}}}\)), \(\kappa _B\) is the
Brillouin zone radius, and $\chi $ is the angle between the x- and the \pmb{ \textit{ s}}-\(\pmb{\text{\textit{$\pmb{s}'$}}}\)direction.

The $\kappa $ integral over the Bessel function \(J_p(\kappa  s)\) { }in Eq. (2.13) extends over the Brillouin zone and requires special attention.
The Brillouin zone is assumed circular to allow for a closed solution. In the earlier approach [12] a smooth cutoff lead to an interaction attractive
in the short range and repulsive with a \(s^{-3}\) law. Such interaction does not fit to the DFT results in [8]. A step cutoff produces an oscillating
interaction with a \(s^{-3/2}\) law and convergence problems when performing lattice sums. The way out is an intermediate cutoff and an oscillating
interaction with a \(s^{-3}\) law, approximated by { }

\[V_{\text{\textit{kl}}}(s,\chi ) =(2\pi )^{-1}\sum _p \omega _{\text{\textit{kl}},p} \cos (p \chi )\cos (\text{p$\pi $}/2)2^{-1-p} \kappa _B{}^3
\left(s \kappa _B\right){}^p\Gamma \left(\frac{3+p}{2}\right)*\]

\[\, _pF_q\left(\left(\frac{3+p}{2}\right),\left(\frac{5+p}{2},1+p\right),-\frac{1}{4} s^2 \kappa _B{}^2\right)/\left(1+\left(s\left/s_0\right.\right){}^{3/2}\right),\text{
           }(2.14)\]

where \(\Gamma (p)\) denotes the Gamma function and \(\, _pF_q\)(a,b,z) the generalized hypergeometric function. 

\subsection*{2.4. Eigenvalues for Pd(100)}

Tab.1 shows the dimensionless coefficients \(\hat{\omega }_{\text{\textit{kl}}\text{\textit{$,$}}p}\) for Pd(100) using elastic constants \(c_{11}\)=221,
\(c_{12}\)=171, \(c_{44}\)=70.8 GPa [20].\\

\(\begin{array}{|c|c|c|c|c|c|}
\hline
 \text{\textit{$k$}} & \text{\textit{$l$}} & \text{\textit{$\hat{\omega }_{\text{kl},0}$}} & \text{\textit{$\hat{\omega }_{\text{kl},2}$}} & \text{\textit{$\hat{\omega
}_{\text{kl},4}$}} & \text{\textit{$\hat{\omega }_{\text{kl},8}$}} \\
\hline
 1 & 1 & -0.9431 &   & 0.1009 & 0.0016 \\
\hline
 1 & 2 & -0.66577 & -0.73706 & -0.07129 &   \\
\hline
 2 & 2 & -1.1521 & -1.0424 & 0.1097 &   \\
\hline
 1 & 3 & -0.66577 & 0.73706 & -0.07129 &   \\
\hline
 2 & 3 & 0.21057 & 0. & -0.21057 &   \\
\hline
 3 & 3 & -1.1521 & 1.0424 & 0.1097 &   \\
\hline
\end{array}\)

Table 1. Coefficients \(\hat{\omega }_{\text{\textit{kl}}\text{\textit{$,$}}p}\) for Pd(100).\\

The \textit{ k}=\textit{ l}=1 coefficients belong to a monomer-monomer (pair) interaction, the non vanishing value of { }\(\hat{\omega }_{\text{\textit{$11,4$}}}\)
and \(\hat{\omega }_{\text{\textit{$11,8$}}}\) stem from the elastic anisotropy of the substrate palladium .\\
The \textit{ k}=1, \textit{ l}=2,3 coefficients belong to monomer-dimer (trio) interactions, we note the opposite sign of the \textit{ p}=2 values
and the identical \textit{ p}=0 and \textit{ p}=4 values.\\
The \textit{ k}=2,3, \textit{ l}=2,3 coefficients belong to dimer-dimer (quarto) interactions, we note the opposite sign of the \textit{ p}=2 values
and the identical \textit{ p}=0 and p=4 values for \textit{ k}=\textit{ l}=2 and \textit{ k}=\textit{ l}=3 respectively. { } { }

The above values reflect symmetries on the (100) surface.

\section*{3. Lattice sum }

In the following the elastic energy will be formulated as lattice sums for 3 parts, consisting of\\
- a monomer-monomer part\\
- a dimer-dimer part\\
- a monomer-dimer part\\
which allows to independently inspect the influence of the stress parameters \(P_1\) and \(P_2\) on the elastic energy and to perform the fit to
DFT configuration energies finally. 

\subsection*{3.1. Elastic energy of adatom configurations }

A configuration of \textit{ m} adatom monomers on a discrete lattice of substrate positions $\{$\textup{ \(\pmb{\text{\textit{$s_i$}}}\)}$\}$ is
described by the monomer density

\[\rho _1(\pmb{s})=\sum _{i=1}^m  \delta \left(\pmb{s}\text{\textit{$-$}}\pmb{s_i}\right) n_i\text{                                             
                            }(3.1)\]

with \(n_{\text{\textit{$i$}}}\in \{0,1\}\) denoting occupation numbers of lattice positions \textup{ \(\pmb{s_i}\)} and \textup{ \(\delta \left(\text{\textit{$\pmb{s}-\pmb{s_i}$}}\right)\text{
 }\)}is the Dirac Delta function. The \(n_{\text{\textit{$i$}}}\) should not be confused with the normal vector in section 2. 

While the monomer stress tensor \(\pmb{\text{\textit{$P_1$}}}\) is located at the adatom lattice points, the dimer stress tensors\(\text{  }\pmb{\text{\textit{$P_2$}}}\)
and \(\pmb{\text{\textit{$P_3$}}}\) are located between two nearest neighbor adatoms. The density of an x-directed adatom dimer configuration of
type \textit{ k=2} with occupation numbers \textup{ \(n_i\)} and \textup{ \(n_{\text{\textit{ix}}}\)} and positions at the center of { }nearest neighbor
positions\textit{  }$\{$\pmb{ \textit{ \(s_i\)}}$\}$ and $\{$\pmb{ \textit{ \(s_{\text{ix}}\)}}$\}$ is given by 

\[\rho _2(\pmb{\text{\textit{$s$}}})=\sum _{i=1}^m  \delta \left(\pmb{s}- \pmb{\text{\textit{$s_i$}}}\text{\textit{$-$}}\pmb{\Delta }_x\right)\text{
 }n_i n_{\text{\textit{ix}}} .\text{                                                              }(3.2)\]

Here we have defined \textup{ \(\pmb{\Delta }_x\)}=\textup{ \(\left(\pmb{\text{\textit{$s_{\text{ix}}$}}}- \pmb{\text{\textit{$s_i$}}}\right.\)})/2.
Likewise the density of an y-directed adatom dimer configuration of type \textit{ k=3} with positions at the center of nearest neighbor positions\textit{
 }$\{$\pmb{ \textit{ \(s_i\)}}$\}$ and $\{$\pmb{ \textit{ \(s_{\text{iy}}\)}}$\}$ is given by 

\[\rho _3(\pmb{\text{\textit{$s$}}})=\sum _{i=1}^m  \delta \left(\pmb{s}- \pmb{\text{\textit{$s_i$}}}\text{\textit{$-$}}\pmb{\Delta }_y\right)\text{
 }n_i n_{\text{\textit{iy}}} .\text{                                                              }(3.3)\]

Here we have defined \textup{ \(\pmb{\Delta }_y\)}=\textup{ \(\left(\pmb{\text{\textit{$s_{\text{iy}}$}}}- \pmb{\text{\textit{$s_i$}}}\right.\)})/2.\\
With Eqs. (3.2), (3.3) and (2.7) the elastic energy of the \textit{ m} adatom configuration reads

\[H_{\text{el}}=H_{11}+H_{12} +H_{13}+H_{22}+H_{23} +H_{33} ,\text{     }(3.4)\]

with

\[H_{11}=\sum _{i=1}^m  \sum _{j=2}^i V_{11}\left(\pmb{\text{\textit{$s_i$}}} ,\pmb{\text{\textit{$s_j$}}}\right)n_i n_j\text{                  
                    }(3.4.a)\]

\[H_{12}=\sum _{i=1}^m \sum _{j=1}^m V_{12}\left(\pmb{\text{\textit{$s_i$}}} ,\pmb{\text{\textit{$s_j$}}}+\pmb{\Delta }_x\right)n_i n_j n_{\text{\textit{jx}}}\text{
                    }(3.4.b)\]

\[H_{13}=\sum _{i=1}^m  \sum _{j=1}^m V_{13}\left(\pmb{\text{\textit{$s_i$}}} ,\pmb{\text{\textit{$s_j$}}}+\pmb{\Delta }_y\right)n_i n_j n_{\text{\textit{jy}}}\text{
                   }(3.4.c)\]

\[H_{22}=\sum _{i=1}^m  \sum _{j=1}^i V_{22}\left(\pmb{\text{\textit{$s_i$}}} +\pmb{\Delta }_x,\pmb{\text{\textit{$s_j$}}}+\pmb{\Delta }_x\right)n_i
n_{\text{\textit{ix}}}n_jn_{\text{\textit{jx}}}\text{               }(3.4.d)\]

\[H_{23}=\sum _{i=1}^m  \sum _{j=1}^m V_{23}\left(\pmb{\text{\textit{$s_i$}}}+\pmb{\Delta }_x ,\pmb{\text{\textit{$s_j$}}}+\pmb{\Delta }_y\right)n_i
n_{\text{\textit{ix}}}n_j n_{\text{\textit{jx}}}\text{                 }(3.4.e)\]

\[H_{33}=\sum _{i=1}^m  \sum _{j=1}^i V_{33}\left(\pmb{\text{\textit{$s_i$}}}+\pmb{\Delta }_y ,\pmb{\text{\textit{$s_j$}}}+\pmb{\Delta }_y\right)n_in_{\text{\textit{iy}}}
n_j n_{\text{\textit{jy}}}\text{  }.\text{              }(3.4.f)\]

Comparing (3.4) with the Hamiltonian used in [8] we note\\
- the elastic expansion (3.4) comprises pair-\textup{ \(\text{  }V_{11}\)} , trio- \textup{ \(V_{1x}\)} , quarto- \textup{ \(V_{\text{xy}}\)} { }interactions
but no higher ones\\
- the { }\textup{ \(V_{kl}\left(\pmb{\text{\textit{$s_i$}}} ,\pmb{\text{\textit{$s_j$}}}\right)\)} are the equivalents to the lateral interactions
in [8].

Eq. (2.12) allows to regroup Eq. (3.4) as sum of elements depending on the parameters \(P_1\) and \(P_2\) for fitting purposes 

\[H_{\text{el}}=P_1P_1 h_{11} +P_1P_2 h_{12}+P_2P_2h_{22}.\text{                             }(3.5)\]

The coverage dependent binding energy per adatom\textup{ \(E_b\)} is the sum of the binding energy of an isolated adatom \(E_b^{\text{on}-\text{site}}\)
{ }and of its interaction energy, in the present model 

\[E_b=E_b^{\text{on}-\text{site}} + \left.H_{\text{el}}\right/m .\text{              }(3.6)\]

\begin{doublespace}
\noindent\(\pmb{ }\)
\end{doublespace}

\textup{ \(E_b^{\text{on}-\text{site}}\)} in the present model contains the self term \textup{ \(V_{11}\left(\pmb{\text{\textit{$s_i$}}} ,\pmb{\text{\textit{$s_i$}}}\right)\)},
describing the strain energy of an isolated adatom. 

\subsection*{3.2. Adatom configurations on Pd(100)}

The lattice sums in Eq. (3.4) have been evaluated for the 5 hollow site binding energies as published in [8]. For the elastic model the stress parameters
\textup{ \(P_1\)} and \textup{ \(P_2\)} together with the Brillouin zone limit \(\kappa _B\) have been fitted to the known binding energies. The
sums have been truncated for elements with \textup{ \(\text{Abs}\left(\pmb{\text{\textit{$s_i$}}} -\pmb{\text{\textit{$s_j$}}}\pmb{\text{\textit{$)$}}}\right.>\)}
25\(s_0\).

The DFT calculated binding energies of [8] and the fitted binding energies for \textup{ \(P_1\)}=11.626 meV, \textup{ \(P_2\)}=-1.9 meV, and for
\(\kappa _B\)=6.116 \(s_0^{-1}\), are compared in Tab. 2. The numerical values for \textup{ \(P_1\)} and \textup{ \(P_2\)} were derived under the
convenience assumption \(c_{44}\)\(s_0^3\)=1 meV for allowing a direct evaluation of Eq. (2.14). The elastic model seems to reasonably interpret
the few binding energies available. { }\(E_b^{\text{on}-\text{site}}\) could not be calculated directly in DFT; its value in [8] was chosen identical
to the DFT binding energy at 1/9 coverage, assuming zero interaction for configurations with smaller coverage. The elastic model, of course, shows
a lower on-site energy.\\

\(\begin{array}{|c|c|c|c|}
\hline
 \text{binding} \text{energy}/\text{meV} & \text{coverage}/\text{ML} & \text{DFT} & \text{elastic} \text{model} \\
\hline
 E_b^{\text{on}-\text{site}} & 0 & -1249 & -1310 \\
\hline
 (3\text{x3})-O & 1/9 & -1249 & -1259 \\
\hline
 p(2\text{x2})-O & 1/4 & -1348 & -1341 \\
\hline
 c(2\text{x2})-O & 1/2 & -1069 & -1066 \\
\hline
 (2\text{x2})-3O & 3/4 & -643 & -692 \\
\hline
 (1\text{x1})-O & 1 & -344 & -343 \\
\hline
\end{array}\)

Table 2. Binding energies per adatom in meV units for O-Pd(100) configurations calculated by DFT [8] and using the present elastic model. \\
$\quad $

The elastic model with the above parameter set can now be used to derive further features of the O-Pd(100) system. In Tab.3 the pair interactions
from [8] are compared with those of the present elastic Hamiltonian up to the 9th nearest neighbors. The optimum set of [8] truncates interactions
beyond the 5th nearest neighbors while in the elastic model the interactions range to infinity - almost with decreasing magnitude. The values of
the four nearest neighbor interactions are in consequence equal in sign but different in value. We also note the variances in the value sets of [8]
due to ill-conditioned configuration matrices. The oscillation wavelength for the \textit{ p}=0 term in Eq. (2.14) with \(\kappa _B\) as found in
the fit is close to the lattice constant \(s_0\).\\

\(\begin{array}{|c|c|c|}
\hline
 \text{pair} \text{interaction}/\text{meV} & \text{optimum} \text{set} [8] & \text{elastic} \text{model} \\
\hline
 n.n. & 292 & 228 \\
\hline
 2\text{nd}\text{  }n.n. & 90 & 66 \\
\hline
 3\text{rd}\text{  }n.n. & -50 & -43 \\
\hline
 4\text{th}\text{  }n.n. & -10 & -15 \\
\hline
 5\text{th}\text{  }n.n. & 0 & 20 \\
\hline
 6\text{th}\text{  }n.n. & 0 & 7 \\
\hline
 7\text{th}\text{  }n.n. & 0 & -6 \\
\hline
 8\text{th}\text{  }n.n. & 0 & -9 \\
\hline
 9\text{th} n.n. & 0 & 8 \\
\hline
\end{array}\)

Table 3. Pair interactions in meV units for O-Pd(100) nearest neighbors, taken from the optimum set in [8] and using the present elastic model. \\
$\quad $

 \(\,\)The generalized hypergeometric function \(\, _pF_q\)(a,b,z) in Eq. (2.14) is oscillating as function of z. The \textit{ p}=0 pair interaction
term in Eq. (2.14) with the above parameter set is \\
- repulsive in the 0.6 to 1.1 \(s_0\) and the 1.7 to 2.2 \(s_0\) regions\\
- attractive in the $<$0.6 \(s_0\), 1.2 to 1.6 \(s_0\) and 2.2 to 2.6\(s_0\) regions.

Trio- and quarto interactions appear relevant close to half coverage and beyond when nearest neighbor adatom dimers show up. They are relevant for
interpreting the high coverage binding energies and for the calculation of 3- and 4 adatom clusters containing nearest neighbor dimers. Table 4 shows
the elastic interaction energy per adatom for some of those clusters. The trio terms stem from the short range interaction created by the strain
of isolated adatoms with the strain of pairs. The quarto term of the first 3-adatom cluster stems from the strain created by the pairs in line; in
the case of rectangular pairs the quarto term is negligible. The quarto terms of { }4-adatom clusters differ in sign because in the first case the
pairs are parallel (with attractive interactions) while in the second case 3 adatoms stand in line (with repulsive interaction). Such cluster energies
provide a test case for DFT calculations. Overall we note that the many-body terms are small compared with the pair terms, a consequence of the small
value of \textup{ \(P_2\)}.\\

\(\begin{array}{|c|c|c|c|c|}
\hline
 \text{configuration} & \text{pair} & \text{trio} & \text{quarto} & \text{sum} \\
\hline
 \left(\bar{1},0\right)(0,0)(1,0) & 138 & -5 & 6 & 139 \\
\hline
 (1,0)(0,0)(0,1) & 174 & -2 & 0 & 172 \\
\hline
 \left(\bar{1},1\right)(0,0)(1,0) & 93 & 2 & 0 & 95 \\
\hline
 (0,1)(0,0)(2,0) & 57 & 1 & 0 & 58 \\
\hline
 (0,1)(0,0)(1,0)(1,1) & 261 & -5 & -1 & 255 \\
\hline
 \left(\bar{1},0\right)(0,0)(1,0)(0,1) & 193 & -5 & 5 & 193 \\
\hline
 \left(\bar{1},\bar{1}\right)\left(\bar{1},1\right)\left(1,\bar{1}\right)(1,1) & -33 & 0 & 0 & -33 \\
\hline
 \left(\bar{1},\bar{1}\right)\left(\bar{1},1\right)(1,0)(1,2) & -35 & 0 & 0 & -35 \\
\hline
\end{array}\)

Table 4. Elastic interaction energy per adatom in meV units for oxygen clusters on Pd(100). Configurations are indicated by primitive cell coordinates;
pair-, trio- and quarto contributions are shown with their sum.\\

The two last lines of Table 4 also shows the interaction energy of some 4-adatom clusters with further distant adatoms. It is interesting to note
that the \(\left(\bar{1},\bar{1}\right)\left(\bar{1},1\right)\left(1,\bar{1}\right)(1,1)\) { }squares { }and the \(\left(\bar{1},\bar{1}\right)\left(\bar{1},1\right)(1,0)(1,2)\)
parallelograms show almost the same elastic energy per adatom; this comes from the 4th n.n. and 8th n.n. attraction and the 5th n.n. repulsion (cf.
Table 3). In consequence a p(2x2) adatom lattice appears unlikely.

\section*{4. Discussion}

In this section limitations and applicability and of the model are reviewed, the results of numerical calculations are discussed and aspects for
further research are sketched. { } { } 

\subsection*{4.1. Restrictions and limitations of the model}

The model is based on the theory of elasticity in the substrate and on the lateral stress adatoms apply to the surface. Key assumption is the mechanism
by which adatoms and dimers interact. Monomer adatoms sitting on fourfold adatom sites expand or contract the substrate by creating isotropic stress.
Trio- and quarto interactions are due to the anisotropic stress adatom pairs exert to the substrate when they are bound electronically and stretched
(or compressed) due to their position on substrate sites.

The electronic interaction of adatoms is assumed restricted to adjacent substrate atoms. The unknown electronic binding energy of adatom pairs is
ignored in the present model. Adatom dipole-dipole interactions may also play a role. 

An elastic continuum model for the substrate may be rated inadequate for describing short range effects, but describes well long range effects and
elastic anisotropies (strong in case of adatom dimers). The cutoff mechanism used in Eq. (2.14) avoids a { }\(s^{-3}\) singularity and therefore
extends the models reach towards small distances. Assuming a circular Brillouin zone is a risk, so Eq. (2.14) can only be an approximation. The search
for a most realistic cutoff function is currently left open.

The restriction to high symmetry adatom locations has the advantage of stress parameter \(P_k\) degeneracy. On (100) surfaces \(P_3\)=\(P_2\) due
to the equivalence of x-directed and y-directed dimers. This reduces the number of free model parameters. The dependence of stress parameter values
\(P_k\) from coverage could be an issue. Its value for isolated dimers could differ from its value in islands since the bond between adatoms depends
on their coordination.

A further key assumption is an ideal flat surface, i.e. the absence of steps which are known for their significant attractive or repulsive interaction
with adatoms. 

The present model parameters rely on the DFT calculations in [8]. The adatom configuration base and their DFT energies used in this paper are restricted
to the 5 configurations published and thus are not at all exhaustive, but allow a first glance on their elastic energy. Utilization of the full range
of the DFT energy results in [8] would allow a consistency proof. The methods used to derive adatom interactions and their near order should remain
valid also in case of a broader base. 

The existence of long range elastic interactions would imply significant influence of the super-cell size used in the DFT calculations used in [8].
Variations of the slab size could prove size effects to the configuration energies.

\subsection*{4.2. Calculation results}

The model of strain mediated interacting oxygen adatoms turns out to interpret well the binding energies of O on Pd(100) calculated with DFT methods.
The few free model parameter support the validity of the elastic model. In consequence the concept of a lattice-gas Hamiltonian without considering
the possibility of long range interactions appears questionable.

Tab. 3 shows only a few nearest neighbor elastic model pair interactions, many more neighbors have been used when calculating the binding energies
shown in Tab. 2. 

The results for the interaction of dimers with monomers or dimers in Tab. 4 show smaller values for interactions than those in [8]. Such reduced
values could avoid the \(T_c\)($\theta $) anomaly obtained with large interaction values in [8]. We note that in [8] only those adatom clusters show
significant trio- and quarto interaction values which contain nearest neighbor dimers - in line with the current elastic model. 

\subsection*{4.3. Open questions and further aspects}

The search for a cutoff function between a smooth exponential and a hard Heaviside function could lead to an oscillating interaction with a proper
decay. { } 

DFT calculations of 2, 3, 4 adatom clusters could provide insights complementary to calculations of ordered adatom structures. They could directly
be compared with the elastic energies of Tab. 4 and act as proof.

A theoretical (DFT) model to determine the magnitude of the stress parameters \(P_k\) could determine the elastic adatom interaction directly.

The DFT calculations [8] using a finite slab should be varied in size to cover long reach effects. 

Other substrate materials than Pd with different elastic constants and different types of adatoms may be interesting to compare the theory with a
broader range of experimental findings. 

The interaction of monomers and dimers consisting of different adatom species is in principle covered by the elastic theory, simulations could help
to assess its applicability.

\section*{5. Summary}

The binding energies of five different O-Pd(100) configurations, calculated by first principles methods [8], have been successfully interpreted by
a Hamiltonian of substrate strain mediated interactions comprising pair-, trio-, and quarto terms with only 3 free parameters. It is concluded that
the often used CE method of building a Hamiltonian without considering longer range interactions is dangerous and can be misleading. A key feature
of the interaction alternative is its analytic nature, allowing statistical methods different from Monte Carlo simulations to derive surface order.
Limitations of the model are analyzed and open questions are addressed. 

\section*{References}

[1] T.L.Einstein, Interactions between Adsorbate Particles, in: Physical Structure of Solid Surfaces (W.N.Unertl ed.), Elsevier, Amsterdam (1996)

[2] J.V.Barth, G.Costantini, K.Kern, Nature 437, 671 (2005)

[3] H.Brune, Creating Metal Nanostructures at Metal Surfaces Using Growth Kinetics, in: Handbook of Surface Science Vol.3 (E.Hasselbrink and B.I.Lundqvist
ed.), Elsevier, Amsterdam (2008) 

[4] H.Ibach, Surf.Sci.Rep. 29, 195 (1997)

[5] F.Komori, S.Ohno, K.Nakatsuji, Progr. Surf. Sci. 77, 1 (2004)

[6] R.I.Masel, Principles of Adsorption and Reaction on Solid Surfaces, Wiley, New York (1996)

[7] K.Reuter, C.Stampfl, M.Scheffler, Ab Initio Atomistic Thermodynamics and Statistical Mechanics of Surface Properties and Functions. In: Handbook
of Materials Modelling, Part A. Methods, (Ed.) S.Yip, Springer, Berlin (2005) 

[8] Y.Zhang, V.Blum, K.Reuter, Phys. Rev. B 75, 235406 (2007) 

[9] M.Todorova, E.Lundgren, V.Blum, A.Mikkelsen, S.Gray, J.Gustafson, M.Borg, J.Rogal, K.Reuter, J.N.Andersen, M.Scheffler, Surf. Sci. 541, 101 (2003)

[10] W.-X.Li, C.Stampfl, M.Scheffler, Phys.Rev.B 65, 075407 (2002)

[11] K.H.Lau, W. Kohn, 1977, Surface Sci. 65, 607 (1977) { } 

[12] W.Kappus, Z.Physik B 29, 239 (1978) 

[13] K.H.Lau, Sol. State Commun. 28, 757 (1978)

[14] H.Horner, H.Wagner, Adv. in Phys. 23,587 (1974)

[15] P.M{\" u}ller, A.Saul, Surf. Sci. Rep. 54, 157 (2004)

[16] W.Kappus, arXiv 1610.07326

[20] A.G.Every, A.K.McCurdy: Table 3. Cubic system. Elements. D.F.Nelson(ed.), SpringerMaterials- \\
The Landolt-B{\" o}rnstein Database (2012)

\section*{Acknowledgement}

This work is dedicated to open minded CE experts.

$\copyright $ Wolfgang Kappus (2017)

\end{document}